\documentclass[sigconf,natbib=true]{acmart}
\usepackage{multirow}
\usepackage{makecell}
\usepackage{enumitem}
\usepackage[ruled,linesnumbered]{algorithm2e}
\usepackage{soul}
\usepackage{tablefootnote}

\newcommand{\PreserveBackslash}[1]{\let\temp=\\#1\let\\=\temp}
\newcolumntype{C}[1]{>{\PreserveBackslash\centering}p{#1}}
\newcolumntype{R}[1]{>{\PreserveBackslash\raggedleft}p{#1}}
\newcolumntype{L}[1]{>{\PreserveBackslash\raggedright}p{#1}}

\AtBeginDocument{%
  }

\copyrightyear{2023} 
\acmYear{2023} 
\setcopyright{acmlicensed}
\acmConference[XXX '23]{Proceedings of XXXX XXXX}{XX XXXX, 2023}{XXXX, XXX, XXX} 
\acmBooktitle{Proceedings of XXXXXX XXXXXX, XXX XXXX, 2023, XXXX, XXXX, XXX} 
\acmPrice{15.00} 
\acmDOI{XXXXX XXXXXX} 
\acmISBN{XXX XXX}

\begin{document}


\title{
  AdaS\&S: a One-Shot Supernet Approach for Automatic Embedding Size Search in 
  Deep Recommender System
}


\author{He Wei}
\orcid{0000-0002-1248-1804}
\authornote{Both authors contributed equally to this research.}
\affiliation{%
   \institution{Machine learning platform department, Tencent TEG.}
   \city{Beijing}
   \country{China}}
\email{whywei@tencent.com}

\author{Yuekui Yang}
\authornotemark[1]
\orcid{0000-0002-8709-3128}
\authornote{Corresponding author.}
\affiliation{%
   \institution{Machine learning platform department, Tencent TEG. And, Department of Computer Science and Technology, Tsinghua University.
   }
   \city{Beijing}
   \country{China}}
\email{yuekuiyang@tencent.com}

\author{Yang Zhang}
\orcid{0009-0000-0698-8361}
\affiliation{%
   \institution{Machine learning platform department, Tencent TEG.}
   \city{Beijing}
   \country{China}}
\email{yizhizhang@tencent.com}

\author{Haiyang Wu}
\orcid{0000-0001-7314-7618}
\affiliation{%
   \institution{Machine learning platform department, Tencent TEG.}
   \city{Beijing}
   \country{China}}
\email{gavinwu@tencent.com}

\author{Meixi Liu}
\orcid{0000-0001-7488-8143}
\affiliation{%
   \institution{Machine learning platform department, Tencent TEG.}
   \city{Beijing}
   \country{China}}
\email{meixiliu@tencent.com}

\author{Shaoping Ma}
\orcid{0000-0002-8762-8268}
\affiliation{%
   \institution{Department of Computer Science and Technology, Tsinghua University. }
   \city{Beijing}
   \country{China}}
\email{msp@tsinghua.edu.cn}

\renewcommand{\shortauthors}{Wei, Yang et al.}

\renewcommand{\shorttitle}{
  AdaS\&S: a One-Shot Supernet Approach for Automatic Embedding Size Search \\
  in Deep Recommender System
}


\begin{abstract}

Deep Learning Recommendation Model(DLRM)s utilize the embedding layer 
to represent various categorical features. 
Traditional DLRMs adopt unified embedding size for all features, 
leading to suboptimal performance and redundant parameters. 
Thus, lots of Automatic Embedding size Search (AES) works focus on 
obtaining mixed embedding sizes with strong model performance. 
However, previous AES works can hardly address several challenges together: 
(1) The search results of embedding sizes are unstable; 
(2) Recommendation effect with AES results is unsatisfactory; 
(3) Memory cost of embeddings is uncontrollable. 
To address these challenges, we propose a novel one-shot AES framework called 
AdaS\&S, 
in which a supernet encompassing various candidate embeddings is built and 
AES is performed as searching network architectures within it. 
Our framework contains two main stages:
In the first stage, 
we decouple training parameters from searching embedding sizes, 
and propose the Adaptive Sampling method to yield a well-trained supernet, which 
further helps to produce stable AES results.
In the second stage, to obtain embedding sizes that benefits the model effect, 
we design a reinforcement learning search process which utilizes the supernet trained 
previously. 
Meanwhile, to adapt searching to specific resource constraint, 
we introduce the resource competition penalty to balance the 
model effectiveness and memory cost of embeddings. 
We conduct extensive experiments on public datasets to show the superiority of AdaS\&S. 
Our method could improve AUC by about 0.3\% while saving about $20\%$ of model parameters. 
Empirical analysis also shows that 
the stability of searching results in AdaS\&S significantly exceeds other methods.

\end{abstract}

\ccsdesc[300]{Information systems~Recommender systems}

\keywords{
  Recommendation, Embedding size, Supernet, Reinforcement Learning
}

\maketitle



\section{INTRODUCTION}
\label{sec.intro}

Deep Learning Recommender Model(DLRM)s often take a large amount of categorical 
features as input, and utilize a embedding layer to convert them to 
low-dimensional embedding vectors \cite{qu2018product}. 
Consequently, the embedding layer plays an important role since 
it dominates the number of parameters as well as 
the effect of model prediction \cite{autodim,yan2021learning}.
Traditional DLRMs generally assign a unified embedding size (abbreviated as ``Emb-size''
 in this work). 
However, unified Emb-size ignores the heterogeneity of features and 
suffers from following issues: (1) 
Inferior performance: setting a unified Emb-size could impair the expressing 
of feature fields with large cardinality (number of unique feature values), while causing 
over-fitting for those with low cardinality.
(2) Memory inefficiency: 
Unified Emb-size could 
cause redundant embedding parameters as well as unnecessary computation overhead, 
especially for features contribute less to final prediction.
Therefore, different Emb-sizes is highly desired now and become 
a focal point in researches, 
especially for online recommendation where inference speed and 
memory cost become the bottleneck.


We denote the task of obtaining different Emb-size
for features as AES (Automatic Embedding size Search). 
Existing AES methods can be categorized as heuristic, pruning and NAS-based methods. 
Heuristic methods tend to allocate Emb-sizes with pre-defined rules.
For example, MDE \cite{ginart2021mixed} utilizes the 
popularity of features to determine the Emb-sizes, but 
its capability is limited by the degeneration of model performance \cite{yan2021learning}. 
As an alternative to heuristic methods, pruning 
methods \cite{liu2021learnable,yan2021learning,qu2022single,kong2022autosrh} reduce Emb-sizes 
from a pre-defined largest dimension according to learnable threshold or mask parameters. 
For example, AutoSrh \cite{kong2022autosrh} use soft selection weights to identify the importance of each
dimension. 
However, the interaction of 
thresholds (masks) and embeddings undermines the stability of their searching outcomes, 
and the complex relation between dimensions is not well-studied.
Recent NAS-based approaches \cite{autodim,zhaok2021autoemb,liu2020automated,zhao2021ameir} view AES as 
searching candidate network architectures, they utilize NAS techniques to 
achieve better performance. Typically, AutoDim \cite{autodim} introduces additional weights 
for embeddings of different Emb-sizes into the network, and optimize them with DARTS \cite{ref18}. 
However, parameter training and dimension searching in these 
works are usually coupled, leading to several weaknesses: (1) joint optimization 
of training and searching could introduce bias during gradient descent and mislead the 
AES to unstable results. (2) the number of embedding parameters 
is not guaranteed, raising the risk of resource costs for model deployment. 
To sum up, though recent pruning and NAS-based methods outperform traditional methods 
in model effectiveness, 
they still suffer from the instability of searched results, and it is also challenging to adapt to varying resource 
constraint (i.e.., memory cost of embedding parameters) while seeking for better Emb-sizes. 


In this paper, we propose a novel approach called AdaS\&S 
(\textbf{Ada}ptive \textbf{S}upernet and RL-\textbf{S}earch) to address the aforementioned 
challenges.
A large network containing all candidate 
embeddings is built (known as ``supernet''),
and Emb-sizes are searched as ``subnet'' architecture, i.e., 
choosing particular embedding tables. 
To be specific, 
(1) \textit{How to generate stable results?}
The decoupled training and searching eliminates the bias in joint optimization \cite{guo2020single}, 
helps to obtain stable outcomes. 
Meanwhile, a supernet with robust parameters could alleviate the uncertainty in 
searching.
(2) \textit{How to obtain better performance?}
We propose the Adaptive Sampling to 
improve the consistency of the trained supernet, so that subnets 
inheriting weights from it could be quite predictive for stand-alone trained ones. 
A Reinforcement Learning (RL) based searching strategy is further 
designed to determine the optimal allocation.
(3) \textit{How to adapt Emb-sizes to resource constraint 
during searching?}
We utilize a regularization 
of total resource cost to guide the RL agent, so that it produces controllable Emb-sizes.

Our approach is experimented on several real-world datasets and shows significant 
advantages. 
Besides, the searching can repeat many times given various 
resource constraints once the supernet is trained, which is 
especially suitable for the dynamic environment of online recommendation.
In summary, the main contribution of this paper are as below:
\begin{itemize}[%
  itemsep=0pt, partopsep=0pt, 
  parsep=\parskip, topsep=0pt,
  leftmargin=1em, listparindent=0em]
\item 
We firstly summarize three main goals for an ideal AES method: stable searching results, 
better recommendation performance with searched Emb-sizes, and being adaptive to specific 
resource constraint.  
We propose the AdaS\&S framework to systematically fulfill all these targets. 

\item Our AdaS\&S method models AES by a one-shot two-stage supernet, 
which successfully decouples Emb-size searching from embedding training. 
Adaptive Sampling is proposed to perform the supernet training. 
Also, we design a Reinforcement Learning method to obtain 
the (near-)optimal Emb-size in the searching space. 
A regularization for total resource is introduced to guide the search result under resource constraints.

\item We conduct extensive experiments on several datasets to demonstrate the superiority of 
our approach over other competitive methods. 
Empirical results show that AdaS\&S could successfully 
address mentioned challenges of AES.
\end{itemize}


\section{PRELIMINARY}
\label{sec.preliminary}
\subsection{Features in DLRMs} 
\label{sec.featuresDLRMs}
In real-world large-scale recommendation system 
\cite{zhang2019deep,ref5,ref6,ref7,ref10}, 
there are a lot of 
different feature fields, 
in which categorical features account 
for a large portion. 
Those features are often stored in 
high-dimensional one-hot or multi-hot vector, 
and are further converted to meaningfully embeddings 
in the latent space. 
Characteristics of these feature fields can be quite different. 
First, the number of distinct feature values (i.e. cardinality) 
in each fields varies a lot. 
%
%
Second, there may be semantic overlapping and value correlation among 
feature fields. 
Last, some feature fields are much more important than 
others in predicting recommendation results.
Therefore, a non-unified assignment of Emb-sizes 
is believed to 
improve the effect and efficiency of DLRM.

\subsection{Base Architecture of DLRM} 
\label{sec.base_arch}
\begin{figure}[h]
  \centering
  \includegraphics[width=0.65\linewidth]{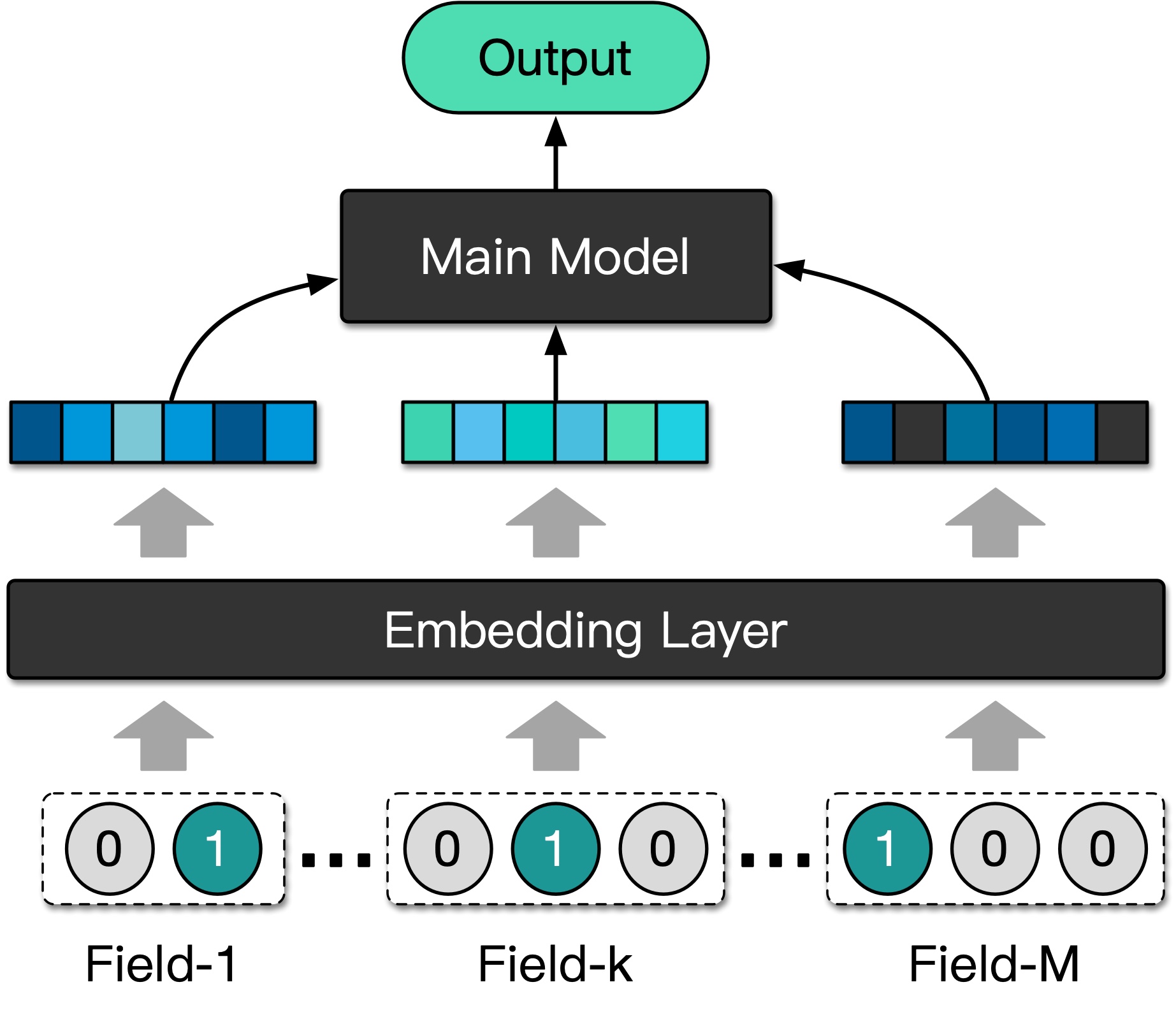}
  \caption{
    Base architecture of the DLRMs. 
  }
  \Description{base architecture}
  \label{fig.base_arch}
  \vspace{-2pt}
\end{figure}

DLRMs 
can be implemented as various popular architectures like 
WDL \cite{ref5}, DeepFM \cite{ref6}, DIN \cite{ref10}. 
There are usually three base components in those models: embedding 
layer, main model, and the output layer, which is shown in
Figure \ref{fig.base_arch}. 

\hspace{-1em}\textbf{Embedding Layer:} 
Assume that there are $M$ categorical feature fields in input data, 
each feature is represented as a binary one-hot or multi-hot encoded 
vector as mentioned above, then the data sample can be formulated as 
$\mathbf{x} = [x_1, x_2, ..., x_M]$, 
where $x_i$ is one-hot or multi-hot vector of the $i$-th field. 
The embedding layer then map these 
$x_i$ to low-dimensional real-valued vectors $e_i = V_i \cdot x_i$,  
where $V_i \in R^{n_i \times d}$ is the embedding space of $i$-th field, 
$d$ is the embedding dimension, $n_i$ is the cardinality of the $i$-th 
field. All parameters in embeddings (a.k.a. embedding table) is 
$\mathbf{V} = \{ V_1, V_2, ..., V_m\}$.
The output of embedding layer is $\mathbf{E}=(e_1, e_2, ..., e_m)$.  

\hspace{-1em}\textbf{Main Model:} 
The main model takes $E$ as input, and employs some 
advanced network architectures like Multi-Layer Perception(MLP) 
\cite{li2019sustainable}, 
Factorization Machine(FM) \cite{originalFM}
and Transformer \cite{vaswani2017attention}
to further process the input information. 
Denoting the parameters of main model as $\Theta$, then the mapping $f$ from 
input $x$ to prediction $\hat{y}$ can be formulated as: 
\begin{flalign}
\begin{split}
\label{eq.yfe}
  \hat{y} = f(x|\mathbf{V},\Theta),
\end{split}
\end{flalign}

\hspace{-1em}\textbf{Output Layer:} 
Typically in CTR prediction, the problem is formulated as a 
binary classification task, 
the output $\hat{y}$ of DLRM is a predict score for the likelihood of click. 
During training, the cross-entropy loss is utilized for updating 
the network:
\begin{flalign}
\begin{split}
\label{eq.loss}
  \mathcal{L}(y, \hat{y};\mathbf{V},\Theta) = -ylog(\hat{y}) - (1-y)log(1-\hat{y}),
\end{split}
\end{flalign}
where $y \in \{0, 1\}$ is the ground-truth label.

\subsection{One-Shot Supernet}
\label{sec.preliminary.one-shot}
Recent NAS approaches adopt weight sharing to avoid the expensive computation. 
All candidate network architectures (i.e., subnet) inherit weights from a 
large network (i.e., supernet). 
The design of decoupling the supernet training from the architecture search is 
known as One-Shot NAS 
\cite{chu2021fairnas,kim2022supernet,chen2021one,you2020greedynas,cai2019once}, 
which often consists of following four steps: 
(1) define a search space and implement it as a supernet. 
(2) train the supernet weights
to make subnets similar to those trained stand-alone. 
(3) search candidate architectures using the pre-trained supernet with 
various strategies.
(4) select the top performing architecture and retrain it.

The key difference of one-shot supernet and DARTS \cite{ref18} is that 
one-shot supernet 
conducts step (2) and (3) in a sequential manner \cite{guo2020single}. 
Since supernet trained in Step (2) guides the searching process in Step (3), 
the evaluation results using inherited supernet weights should be 
predictive for those using stand-alone trained weights. 
To evaluate the quality of trained supernet, 
ranking correlation metric, e.g. Kendall Tau, is employed to 
measure the consistency between one-shot models and stand-alone trained 
ones \cite{yu2019evaluating}. 


\section{FRAMEWORK}
\label{sec.framework}
The main goal of AES in this work is to select the optimal 
Emb-sizes for each feature field to achieve best recommendation performance. 
As in previous works like 
AutoDim \cite{autodim} and 
ESAPN \cite{liu2020automated}, we can formulate this into an 
optimization problem as follows:
\begin{flalign}
\begin{split}
\label{eq.total_aim}
\mathbf{d}^\ast,\mathbf{V}^\ast, \Theta^\ast=\mathop{\arg\min}\limits_{\mathbf{d} \in \mathcal{D}} 
  \mathcal{L}(y, \hat{y}; \mathbf{d}, \mathbf{V}, \Theta),
\end{split}
\end{flalign}
where $\mathbf{d} = [d_{1}, d_{2}, ..., d_{M}]$ is the Emb-size 
assignment result for $M$ features, i.e., $V_i \in R^{n_i \times d_i}$. 
$\mathcal{D}$ is the search space of the Emb-size, 
if there are $M$ features and $T$ candidate Emb-sizes, 
the search space $\mathcal{D}$ will contain $T^{M}$ elements. 
As $M$ and $T$ can be quite large,
it is impossible to traverse all cases by 
Eq.\eqref{eq.total_aim} to obtain the optimal Emb-size assignment 
due to the computation cost. 

\begin{figure}[t!]
  \centering
  \includegraphics[width=1.0\linewidth]{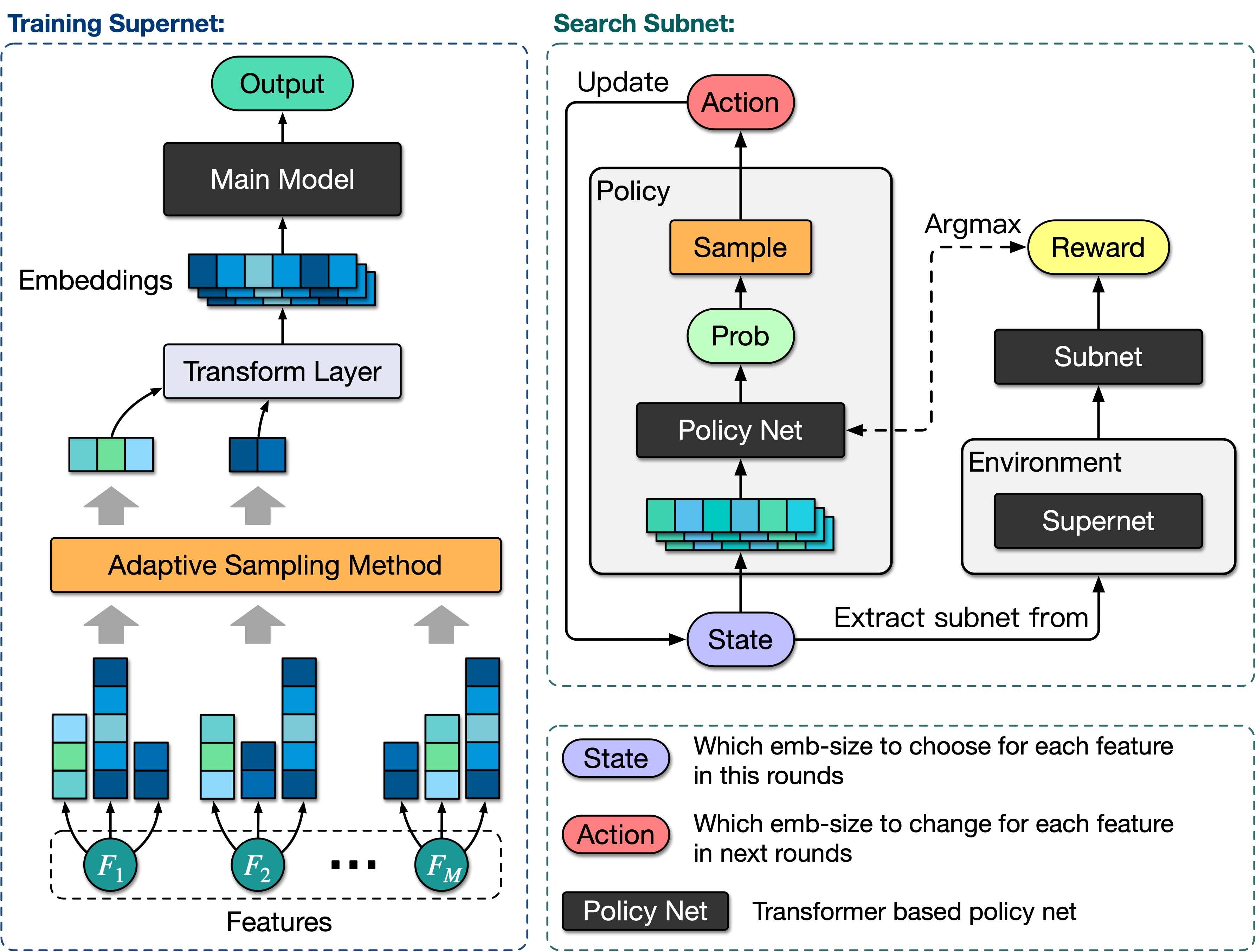}
  \caption{
    Overview of the proposed AdaS\&S framework. 
    The left part is the stage of training the Adaptive Supernet. 
    The right part is the stage of RL-Search process. 
  }
  \Description{overview}
  \label{fig.overview}
  \vspace{-2pt}
\end{figure}

To solve Eq.\eqref{eq.total_aim}, 
we propose the two-stage one-shot framework AdaS\&S. 
It contains 
two main stages: training supernet and searching subnet (see Figure \ref{fig.overview}). 
To be more precise, (1) in the training stage, we train the supernet that encompasses 
all candidate Emb-sizes and aim to make all embeddings fully expressive. We design the 
Adaptive Sampling method to obtain a supernet with informative parameters and 
high consistency. 
(2) in the searching stage, we search the optimal Emb-size with a RL process. A policy network 
serves as RL agent that takes the state (chosen Emb-sizes at current step) and yields the action 
(chosen Emb-sizes for next step) under the resource competition penalty. 
The RL agent keeps adjusting Emb-sizes until this process ends (the searching being converged); 
(3) finally, an additional retraining step is conducted to 
produce a DLRM with different Emb-sizes where only optimal Emb-sizes are taken into consideration.

Table \ref{table.capabilities} gives a comprehensive comparison of our approach with 
others on whether they addressed aforementioned challenges. 
To the best of our knowledge, AdaS\&S is the only method that 
tackles all challenges in design. 
Moreover, our framework could be implemented into various DLRMs, and the one-shot nature 
(supernet training decoupled with Emb-size searching) allows 
multiple times of searching after the training stage, 
which is particularly efficient 
for online applications with ever-changing requirements. 
Next we will present details of several main components 
in AdaS\&S. 
\begin{table}[h!]
\vspace{2pt}
\renewcommand\arraystretch{1.1}
\centering
\caption{
  Capabilities of AES methods. 
  ``Stable'' for obtaining stable search results, 
  ``Performance'' for improving model performance with new Emb-sizes,
  ``Res'' for adapting to the resource constraint of parameters.
}
\label{table.capabilities}
\resizebox{0.85\linewidth}{!}{%
\begin{tabular}{lcccc}
\Xhline{0.85pt}
\multicolumn{1}{c}{Model}  & Category  & Stable  
  & Performance  & Res \\
\Xhline{0.5pt}
MDE \cite{ginart2021mixed}   & Heuristic  & $\boldsymbol{\surd}$   
      & $\boldsymbol{\times}$  & $\boldsymbol{\surd}$ \\
AutoDim \cite{autodim} & NAS  & $\boldsymbol{\times}$   
      & Sub\tablefootnote{``Sub'' indicates sub-optimal performance (than AdaS\&S).}  & $\boldsymbol{\times}$ \\
AutoEmb \cite{zhaok2021autoemb} & NAS  & $\boldsymbol{\times}$   
      & Sub  & $\boldsymbol{\times}$ \\
ESAPN \cite{liu2020automated} & NAS  & $\boldsymbol{\times}$   
      & Sub  & $\boldsymbol{\times}$ \\
AMTL \cite{yan2021learning} & Pruning  & $\boldsymbol{\times}$   
& Sub  & $\boldsymbol{\times}$ \\
PEP \cite{liu2021learnable} & Pruning  & $\boldsymbol{\times}$   
& Sub  & $\boldsymbol{\times}$ \\
SSEDS \cite{qu2022single} & Pruning  & $\boldsymbol{\times}$   
& Sub  & $\boldsymbol{\surd}$ \\
\textbf{AdaS\&S} & NAS (One-Shot)   & $\boldsymbol{\surd}$   
& Better\tablefootnote{``Better'' indicates better performance than others}  & $\boldsymbol{\surd}$ \\
\Xhline{0.8pt}
\end{tabular}%
}
\end{table}

\subsection{Adaptive Supernet}
\label{sec.supernet}


Following the paradigm in Section.\ref{sec.preliminary.one-shot}, 
we propose our
one-shot supernet to encompass all possible subnets (Emb-size allocations) in the 
search space. This enables our method to have the following strengths: 
(1) parameters in embeddings and main model could be decoupled from searching Emb-sizes;
(2) searching can be conducted independently and repeatedly after supernet training. 
Our supernet shall also meet the following requirements:
(1) it should be memory-efficient when handling with larger search space $\mathcal{D}$,
(2) the training of parameters shall be adaptively adjusted in order to 
fully express each candidate embeddings, 
(3) the training stage shall produce a supernet with high consistency to better guide the searching, 
and (4) training supernet shall be easily integrated into various DLRMs.
To meet these requirements, we introduce several techniques into our supernet training, 
including supernet schemes, 
Adaptive Sampling, and unifying different Emb-sizes.
\begin{figure}[h!]
  \centering
  \includegraphics[width=0.95\linewidth]{
    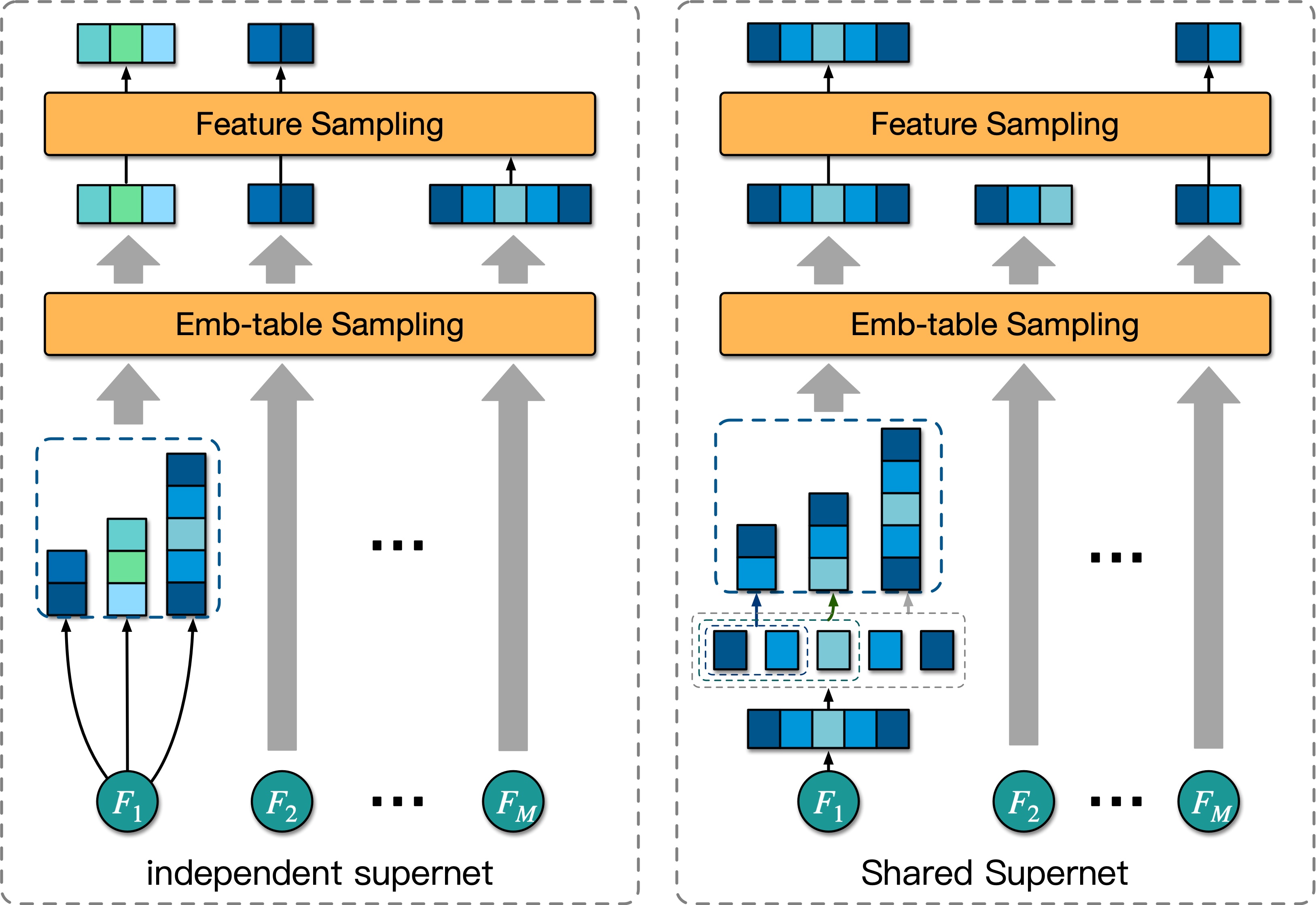
  }
  \caption{
    Two schemes of supernet training.
  }
  \Description{Scheme}
  \label{fig.sscheme}
  \vspace{-2pt}
\end{figure}

\hspace{-1em}\textbf{Supernet Scheme:} 
We devise two different supernet schemes to fulfill candidate embeddings 
with different sizes, namely independent supernet and shared supernet
(Figure \ref{fig.sscheme}). 
In \textit{Independent Scheme}, we initialize several independent embedding tables for 
each candidate Emb-sizes, each individual table serves for one candidate Emb-size. 
This scheme ensures each candidate trained independently, producing more accurate parameters. 
Yet it may place greater demands on hardware during training. 
The number of total embedding parameters in this scheme is $\sum{n_i}\sum_{t=1}^{T}{d_t}$.

While larger search space $\mathcal{D}$ is preferred and memory 
cost of supernet becomes a concern, \textit{Shared Scheme} is a feasible alternative, in which 
a largest embedding table (known as ``max-table'') with largest candidate Emb-size is initialized 
and smaller candidates with Emb-size $d_t$ are parts of it (by selecting first $d_t$ dimensions).
This scheme is more memory-efficient especially for larger $M$, $T$.

\hspace{-1em}\textbf{Adaptive Sampling:} 
Sampling candidate architectures plays a key role in supernet training and 
shows a significant impact on supernet consistency \cite{guo2020single,chu2021fairnas}. 
To determine a proper way to train the supernet, 
we devise our new sampling method from two perspectives: 
On the one hand, to obtain well-trained parameters in different Emb-sizes, 
it is preferable to increase the chance for training under-fitting embedding tables. 
On the other hand, to distinguish the contribution of different features fields, 
features that are more important should fully express 
their information, thus requiring higher training intensity.
Consequently, we propose a two-layer Adaptive Sampling method: 
\begin{flalign}
\hspace{-1cm}
\begin{split}
\label{eq.adas}
  & \ddot{\mathbf{E}} = 
  AdaS(\mathbf{E}, P_{E}, P_{F}) = 
  FS(\dot{\mathbf{E}}, P_{F}), \\
  & \text{Where, } \dot{\mathbf{E}} 
  = ES(\mathbf{E}, P_{E}). 
\end{split}
\hspace{-1cm}
\end{flalign}

In Eq.\eqref{eq.adas}, 
$\mathbf{E} = \{E_{i,j}\}_{M \times T}$ is the embedding 
list of $M$ features and $T$ candidate Emb-sizes, where $E_{i,j} \in R^{d_{j}}$, 
$P_{E} \in R^{M\times T}$ is the sample rate of Emb-sizes, 
$P_{F} \in R^{M}$ is the sample rate of features. The two layers of Adaptive Sampling 
consist of: 
\begin{itemize}[%
  itemsep=0pt, partopsep=0pt, 
  parsep=\parskip, topsep=0pt,
  leftmargin=1em, listparindent=0em]
\item
$ES(\mathbf{E}, P_{E})$ is the first layer 
called embedding table sampling. 
It builds a categorical distribution over embedding list 
$\mathbf{E}$ according to sample rate $P_{E}$. 
Each time it samples a result $\dot{\mathbf{E}} = \{E_{i}\}_{M}$, which is the 
chosen embeddings for $M$ features. For each feature, $ES$ selects 
exactly one embedding from $T$ candidates.
\item
$FS(\dot{\mathbf{E}}, P_{F})$ is the second layer 
called feature sampling. 
It builds a multivariate Bernoulli distribution over 
$\dot{\mathbf{E}}$ according to sample rate $P_{F}$. 
Each time it samples a result 
$\ddot{\mathbf{E}} = \{E_{i}\}_{\tilde{M}}$, where $\tilde{M} \leq M$. 
$\ddot{\mathbf{E}}$ is the final embedding list of 
selected $\tilde{M}$ features. 
In other words, $FS$ would select a subset of all $M$ features to train.
\end{itemize}

Similar to WAST \cite{sokar2022pay}, 
our sample rate $P_{E}$ is adaptively updated 
based on the training status of embeddings, and 
$P_{F}$ is adaptively updated 
based on the importance of features. We formulate the updating of 
$P_{E}$ and $P_{F}$ as below: 
\begin{flalign}
  & P_{E}^{(t+1)} = P_{E}^{(t)} 
  - LN(
    \frac{1}{T}\sum_{j=1}^{T} Var(E_{\cdot, j})
  ), \label{eq.peee}\\
  & P_{F}^{(t+1)} = P_{F}^{(t)} 
  + LN(\frac{\lambda_{FS}}{M\cdot T}
    \parallel \frac{\partial \mathcal{L}}
      {\partial \dot{\mathbf{E}}} \parallel_{1} 
    + \frac{1-\lambda_{FS}}
      {M\cdot T} \parallel \dot{\mathbf{E}} \parallel_{1} 
    ), \label{eq.pfff}
\end{flalign}
$LN$ is the layer normalization, 
$Var(E_{\cdot, j})$ is the value variance of the $j$-th candidate embedding in 
a field, 
 ${\partial \mathcal{L}}/
{\partial \dot{\mathbf{E}}}$ is the gradient for embeddings, 
$\parallel \dot{{\mathbf{E}}} \parallel_{1}$ is the L1 norm of embeddings, and 
$\lambda_{FS}$ is the hyperparameter for balancing the effects of gradients and L1 norms. 
By Eq.\eqref{eq.peee}, embedding table with insufficiently trained parameters 
(lower variance) will get 
more chance to be sampled than others. 
By Eq.\eqref{eq.pfff}, features contribute more to prediction (larger gradients
 or L1 norm) will 
get more chance to be sampled.

\hspace{-1em}\textbf{Unifying Different Emb-sizes:} 
After sampling, we obtain the embedding list $\ddot{\mathbf{E}}$ containing 
mixed Emb-sizes. 
Since some DLRMs like DeepFM \cite{ref6} require 
same dimensions for input embeddings from all feature fields to model feature interactions, 
we need to transform embeddings in $\ddot{\mathbf{E}}$ into same Emb-size in order to feed them into main model:
\begin{flalign}
\begin{split}
  \tilde{\mathbf{E}} = Transform(\ddot{\mathbf{E}}, d_{f}),
  \label{eq.embTransform}
\end{split}
\end{flalign}
where $\tilde{\mathbf{E}} \in R_{M \times d_{f}}$ is the final 
embeddings of $M$ features, $d_{f}$ is the unified size. 
$Transform$ is implemented as MLP layers, and we introduce $T$ MLPs 
to map $T$ candidate Emb-size $d_1, ..., d_T$ to $d_{f}$ respectively. All candidate embeddings 
with same Emb-size across feature fields shall be transformed with the same MLP, which helps to 
reduce parameters and computation overhead during training stage. 
Batch Normalization \cite{ioffe2015batchnorm} is appended to MLPs to stabilize the magnitude of transformed embeddings 
according to \cite{autodim,zhaok2021autoemb}. 
The overall training process of supernet is shown in Algorithm.\ref{algo.supernet}. 


\begin{algorithm}[h!]
  \caption{Supernet training process}
  \label{algo.supernet}
  \LinesNumbered
  \KwIn{
    Features: $x = [x_{1}, ...,x_{M}]$, 
    Labels: $y$, 
    Supernet scheme: $S$,
    Candidate Emb-sizes: $D = [d_{1}, d_{2}, ..., d_{T}]$,
    Final Emb-size: $d_{f}$; 
  }
  \KwOut{the well-trained Supernet;}
  Initialize $P_{E}, P_{F}$;

  \uIf{$S = \text{Independent}$}{
    Initialize several embedding tables of $D$: 
    $\mathbf{E} = \{E_{i,j}\}_{M\times T}$;
  }
  \uElseIf{$S = \text{Shared}$}{
    Initialize the maximum table: $E_{max}$;

    Intercept $E_{max}$ into several tables of $D$: 
    $\mathbf{E} = \{E_{i,j}\}_{M\times T}$; 
  }

  \For{batch = 0; batch < B; batch++}{
    Feed a mini-batch: $\{x, y\}$ of training set to Supernet;

    Embedding lookup from $\mathbf{E}$ based on $x$; 

    Adaptive Sampling: 
    $\ddot{\mathbf{E}} = AdaS(\mathbf{E}, P_{E}, P_{F})$; 

    Transform to unified size:
    $\tilde{\mathbf{E}} = Transform(\ddot{\mathbf{E}}, d_{f})$;

    Update $W$ by descending 
    $\bigtriangledown_{\mathcal{W}} 
      \mathcal{L}_{train}(\hat{y}(\tilde{E}), y)$;

    Update Sample rate: $P_{E}, P_{F}$ 
    by Eq.\eqref{eq.peee} and Eq.\eqref{eq.pfff}; 
  }
\end{algorithm}

\subsection{RL-Search Process}
\label{sec.rlsearch}
After the training stage, we can search the optimal subnet 
based on the parameters of supernet. 
In NAS, 
genetic algorithm, evolution algorithm, RL, etc. \cite{ref17} are commonly 
used for searching the 
architecture, 
and 
we employ RL for AdaS\&S in this stage. To be specific, 
we design a transformer based policy net capturing the 
complex relationships between features 
to conduct the search process. 
The components of our RL-search process can be formulated as: 

\hspace{-1em}\textbf{(1) State: }
$s \in R^{M}$, the state of Emb-size assignment. 

\hspace{-1em}\textbf{(2) Action: }
 $a \in R^{M}$, 
select the Emb-size for each feature in the next round. 

\hspace{-1em}\textbf{(3) Track: }
$\tau_{l} = [s_{0}, a_{0}, s_{1}, a_{1}, ..., s_{l}, a_{l}]$ 
is the state-action track in $l$-th step in RL process. 

\hspace{-1em}\textbf{(4) Policy: }
$\pi_{\theta}(s_{i})$ is a transformer based policy network. 
It takes current \textbf{state} as inputs and generates a state 
transition probability matrix $\mathcal{P} \in R^{M \times T}$, 
where $\mathcal{P}_{i, j}$ is transition probability of the $i$-th 
feature from the current Emb-size to the $j$-th Emb-size. 
When RL-searching is converged, $\mathcal{P}_{i,j}$ will represent the 
probability that the $j$-th Emb-size will be selected for the $i$-th feature. 

\hspace{-1em}\textbf{(5) Reward: }
$\mathcal{R}(s_{i}, a_{i}) = \lambda_{r} \cdot ACC_{val}(\hat{y}, y)$, 
where $ \lambda_{r}$ is the weight of reward, 
$ACC_{val}(\hat{y}, y)$ is the accuracy of the subnet. 

\hspace{-1em}\textbf{(6) Objective function: } 
We define the objective $\mathcal{J}_{val}(a_{i} | \theta)$ based on 
REINFORCE \cite{williams1992simple} and AutoFSS \cite{wei2023automatic}, 
it can be expanded as: 
\begin{flalign}
\begin{split}
\label{eq.objective}
  \mathcal{J}_{val}(s_{i}, a_{i}| \theta) = 
    -\mathbb{E}_{\pi_{\theta}(s_{i})} 
    [\log(p_{\theta}(\tau_{i}))\cdot \mathcal{R}(s_{i-1}, a_{i})] 
    + \mathbb{P},
\end{split}
\end{flalign}
where, $\mathbb{E}_{\pi_{\theta}(a_{i})}[\cdot]$ denotes the expected 
reward of the track $\tau_{i}$. 
$\mathbb{P}$ is resource competition penalty, which is introduced to 
limit the total memory resources occupied by embedding tables. 

The resource competition penalty $\mathbb{P}$ can be formulated as: 
\begin{flalign}
\begin{split}
\label{eq.penalty}
  \mathbb{P} = \lambda_{r} \frac{1}{M}
    \sum_{i=1}^{M}\sum_{j=1}^{T}
      \left(d_{j} \mathcal{P}_{i,j}\right)
    -\lambda_{c} \frac{1}{M}
    \sum_{i=1}^{M}
      \parallel \mathcal{P}_{i} - \frac{1}{T}\parallel_{2},
\end{split}
\end{flalign}
where $\lambda_{r}, \lambda_{c}$ is 
the hyper-parameter for weights of penalties, 
$d_{j}$ is the $j$-th Emb-size in search space, 
$T$ is the number of candidate Emb-sizes. 
The first part of $\mathbb{P}$ is the Emb-size resources penalty, 
which can limit the memory resources. 
The second part is a competition penalty, 
which keeps Emb-size away from a moderate result, i.e., uniformly choosing a candidate size. 
In this way, RL searching can model the competency of features for resources and grant larger Emb-sizes for more promising features.

The RL-search process is illustrated in Figure \ref{fig.overview} (right side).
After searching, we can obtain the optimal Emb-size 
assignment according to transition probability $\mathcal{P}$. 
It is notable that by adjusting $\lambda_{r}$ and $\lambda_{c}$ in the penalty 
$\mathbb{P}$, 
we can balance the effect and resource during searching. 
For example, to force the searching 
to generate Emb-sizes with less embedding parameters, we can increase the 
value of $\lambda_{r}$.

\subsection{Re-train with Different Emb-size}
\label{sec.retrain}

After the searching stage, optimal Emb-size assignment is obtained by 
$\mathcal{P}$. For the $m$-th feature, optimal Emb-size $d_m^{\ast}$ is the $t_m^\ast$-th 
candidate Emb-size, and: $t_m^{\ast} = argmax_{t}(\mathcal{P}_m)$. 
Since the supernet is trained by sampling embeddings of different sizes, a re-training 
of main model with only optimal $E_m^{\ast}$ is desired to eliminate the influence 
of suboptimal Emb-sizes. As in Section.\ref{sec.preliminary}, we obtain unique embedding 
vectors $[e_1, e_2, ..., e_M]$ for features $[x_1, x_2, ..., x_M]$, and concatenate them to 
feed into hidden layers of the main model. 
As in Section.\ref{sec.supernet}, 
we still employ the transform layer to map embeddings of mixed-sizes 
to a unified size. Without redundant embedding parameters, 
the memory consumption and training time would considerably decrease 
in re-training.


\section{EXPERIMENT}
\label{sec.experiment}
In this section, we design experiments to analyze the following research questions: 
\begin{itemize}[%
  itemsep=0pt, partopsep=0pt, 
  parsep=\parskip, topsep=0pt,
  leftmargin=1em, listparindent=0em]
\item \textbf{RQ1}: How does the proposed AdaS\&S perform on AES problem? Could AdaS\&S outperform 
other past AES methods for obtaining optimal recommendation performance?
\item \textbf{RQ2}: Can the proposed AdaS\&S obtain searching results with more stability? 
\item \textbf{RQ3}: Can the proposed AdaS\&S adaptively search for Emb-sizes according to 
specified resource constraints?
\item \textbf{RQ4}: 
Does Adaptive Sampling in the training stage improve the supernet consistency?
\item \textbf{RQ5}: What is the influence of AdaS\&S on the efficiency of searching Emb-sizes and 
model inference?
\end{itemize}

\subsection{Dataset}



We conduct experiments on three real-world datasets: 

\textbf{Avazu}\footnote{\url{https://www.kaggle.com/competitions/avazu-ctr-prediction/data}} 
consists of 40M users' click records on ads over 11 days, 
and each record contains 22 categorical features. 
We chronologically divide this dataset into training/validation/test set 
by 8:1:1. 

\textbf{Movielens-1M (ML-1M)}\footnote{\url{https://grouplens.org/datasets/movielens/1m/}}, 
a dataset for about 1M records of user-movie ratings. 
As in AutoInt \cite{song2019autoint} and PEP \cite{liu2021learnable}, 
we take samples with ratings $> 3$ as positive samples. 
The original data samples contain several feature fields: 
user ID, gender, age, occupation, zip, movieId, year, genres, 
timestamps. We convert timestamp to 2 categorical features, 
namely weekday (indicating whether it is for weekend or not) and 
hour\_in\_day (indicating the time in 24 hour). 
The training/validation/test set is divided by 8:1:1 based on 
timestamp.

\textbf{Criteo}\footnote{\url{https://www.kaggle.com/datasets/mrkmakr/criteo-dataset}}, 
a dataset contains 13 numerical and 26 categorical features. 
Note that all feature fields are anonymized with MD5 encoding. 
We normalize numerical features by transforming a value $v$ as 
in \cite{autodim}, and then bucket them into categorical features. 
We divide the data samples into training/validation/test set 
with 8:1:1 according to the history of user-item interactions.
The summary of these datasets can be found 
in Table \ref{table.data_ifor}.
\begin{table}[h!]
\vspace{2pt}
\renewcommand\arraystretch{1.1}
\centering
\caption{
Overall information of datasets.
}
\label{table.data_ifor}
\resizebox{.8\linewidth}{!}{%
\begin{tabular}{lccc}
\Xhline{0.8pt}
\multicolumn{1}{c}{Data} & Avazu        & ML-1M   & Criteo          \\
\Xhline{0.5pt}
\# Interactions          & 40,349,512   & 994,833 & 45,840,617     \\
\# Feature Fields        & 22           & 10      & 39           \\
\# Feature Values        & 6,481,134    & 15,703  & 7,605,436   \\
Label                    & click or not & Rating  & click or not  \\
\Xhline{0.8pt}
\end{tabular}%
}
\end{table}

\begin{table*}[h]
\renewcommand\arraystretch{1.1}
\centering
\caption{
Overall Performance of AdaS\&S. 
AdaS\&S-R is the resource-first mode. 
AdaS\&S-E is the effect-first mode. 
P-R indicates the percentage of Parameters Reduction compared to UES-32. 
Higher AUC or lower LogLoss mean better performance. \\
``*'' indicates p-value $<0.001$ in significance test 
(two-sided t-test) with the baseline in same order. 
}
\label{table.overall}
\resizebox{.9\linewidth}{!}{%
\begin{tabular}{c|c|l|cc|ccccc|cc}
\Xhline{0.8pt}
\multirow{2}{*}{Dataset}      & 
\multirow{2}{*}{Base Model}   & 
\multicolumn{1}{c|}{\multirow{2}{*}{Metric}} &
\multicolumn{2}{c|}{Baseline} & 
\multicolumn{5}{c|}{Past AES Method}  & 
\multicolumn{2}{c}{Our Framework}  \\ 
\cline{4-12}
& & & UES-32    & UES-24       & MDE     & AutoDim   & 
AutoEmb   & ESAPN    & SSEDS   & AdaS\&S-R   & AdaS\&S-E   \\
\Xhline{0.5pt}
\multirow{6}{*}{Avazu} & 
\multirow{3}{*}{Wide\&Deep}  & 
    AUC     & 0.7811   & 0.7784   & 0.7806   & 0.7821  & 0.7818
            & 0.7810   & 0.7824   & 0.7822   & \textbf{0.7846}*  \\
& & LogLoss & 0.3872   & 0.3915   & 0.3894   & 0.3859  & 0.3864
            & 0.3878   & 0.3861   & 0.3860   & \textbf{0.3842}*  \\
& & P-R     & - -      & 27.31\%  & 23.45\%  & 8.43\%  & 14.13\%
            & 25.51\%  & 24.09\%  & \textbf{32.97\%}   & 21.58\% \\
\cline{2-12}
& \multirow{3}{*}{DeepFM}  & 
    AUC     & 0.7829   & 0.7803   & 0.7813   & 0.7837  & 0.7841 
            & 0.7831   & 0.7837   & 0.7842   & \textbf{0.7866}*  \\
& & LogLoss & 0.3849   & 0.3880   & 0.3865   & 0.3839  & 0.3831
            & 0.3845   & 0.3834   & 0.3828   & \textbf{0.3807}*  \\
& & P-R     & - -      & 26.19\%  & 24.93\%  & 12.17\% & 10.41\%
            & 23.94\%  & 23.27\%  & \textbf{31.12\%}   & 20.83\% \\
\Xhline{0.5pt}
\multirow{6}{*}{ML-1M} & 
\multirow{3}{*}{Wide\&Deep}  & 
    AUC     & 0.7873   & 0.7846   & 0.7868   & 0.7881  & 0.7879
            & 0.7870   & 0.7880   & 0.7882   & \textbf{0.7905}*  \\
& & LogLoss & 0.5568   & 0.5604   & 0.5585   & 0.5547  & 0.5558
            & 0.5567   & 0.5554   & 0.5550   & \textbf{0.5532}*  \\
& & P-R     & - -      & 24.91\%  & 21.93\%  & 6.10\%  & 17.27\%
            & 24.24\%  & 23.57\%  & \textbf{31.71\%}   & 22.71\% \\
\cline{2-12}
& \multirow{3}{*}{DeepFM}  & 
    AUC     & 0.7891   & 0.7867   & 0.7886   & 0.7898  & 0.7904 
            & 0.7893   & 0.7899   & 0.7901   & \textbf{0.7923}*  \\
& & LogLoss & 0.5547   & 0.5572   & 0.5561   & 0.5539  & 0.5530
            & 0.5542   & 0.5532   & 0.5532   & \textbf{0.5513}*  \\
& & P-R     & - -      & 24.54\%  & 23.29\%  & 11.57\% & 8.81\%
            & 22.72\%  & 22.83\%  & \textbf{30.68\%}   & 21.05\% \\
\Xhline{0.5pt}
\multirow{6}{*}{Criteo} & 
\multirow{3}{*}{Wide\&Deep}  & 
    AUC     & 0.7984   & 0.7955   & 0.7977   & 0.7990  & 0.7988 
            & 0.7981   & 0.7992   & 0.7995   & \textbf{0.8017}*  \\
& & LogLoss & 0.4935   & 0.4964   & 0.4944   & 0.4923  & 0.4923 
            & 0.4932   & 0.4920   & 0.4919   & \textbf{0.4901}*  \\
& & P-R     & - -      & 29.07\%  & 28.51\%  & 17.94\% & 16.03\%
            & 26.59\%  & 27.54\%  & \textbf{34.10\%}   & 25.12\% \\
\cline{2-12}
& \multirow{3}{*}{DeepFM}  & 
    AUC     & 0.8013   & 0.7987   & 0.8005   & 0.8031  & 0.8027
            & 0.8011   & 0.8034   & 0.8030   & \textbf{0.8044}*  \\
& & LogLoss & 0.4892   & 0.4918   & 0.4899   & 0.4873  & 0.4878 
            & 0.4893   & 0.4870   & 0.4876   & \textbf{0.4860}*  \\
& & P-R     & - -      & 28.53\%  & 27.14\%  & 14.03\% & 11.40\%
            & 23.48\%  & 25.74\%  & \textbf{32.81\%}   & 23.94\% \\
\Xhline{0.8pt}
\end{tabular}
}
\vspace{-12pt}
\end{table*}

\subsection{Implement Details}
\label{sec.details}
\textbf{Base Model:} 
We compare AdaS\&S with several popular AES methods: 
MDE \cite{ginart2021mixed}, 
AutoDim \cite{autodim}, 
AutoEmb \cite{zhaok2021autoemb}, 
ESAPN \cite{liu2020automated}, 
SSEDS \cite{qu2022single}. 
We also compare our method with the case 
where all embeddings are in unified Emb-size (dubbed as UES). 
For example, UES-32 means all Emb-sizes are set to 32. 
UES could set a baseline for model performance and parameter number. 
We conduct AES and UES settings 
based on popular DLRM architectures, namely Wide\&Deep \cite{ref5} and DeepFM \cite{ref6}. 
Note that 
the original AutoEmb and ESAPN would search Emb-sizes for each unique feature value, which 
is hardware-unfriendly \cite{lyu2022optembed} for industrial-level applications. 
In our experiments, we employ the main framework of these two methods and 
conduct them in a field-wise manner. 
We employ the original MDE since it is a typical heuristic method which 
determines Emb-sizes by the popularity of feature values.

\hspace{-1em}\textbf{Hyper-parameter:} 
The hyper-parameters in experiments are as below: 
\textbf{(a)} Batch size is 512 for three datasets. 
\textbf{(b)} Learning rate: 
0.001 for Main Model (Supernet training stage), and 
0.0005 for searching stage. 
\textbf{(c)} MLP layer: 
Wide\&Deep and DeepFM use a 3-layer MLP network with 
ReLU as the activation function. 
The width of MLP layer is set to $\{128, 64, 1\}$ for Avazu and ML-1M, 
and $\{256, 128, 1\}$ for Criteo. 
\textbf{(d)} Search space: $D=\{2, 8, 16, 32, 64\}$. 
The largest candidate Emb-size is 64. 
\textbf{(e)} Settings of AdaS\&S: 
We use independent supernet scheme for ML-1M, 
shared supernet scheme for Criteo and Avazu. 
In feature sampling process, $\lambda_{FS} = 0.6$. 
In resource competition penalty, $\lambda_{r} = 0.0025$ and 
$\lambda_{c} = 0.08$ for effect-first mode, 
$\lambda_{r} = 0.005$ and $\lambda_{c} = 0.04$ for resource-first mode.

\hspace{-1em}\textbf{Metric:} Following DeepFM \cite{ref6}, 
we use the commonly-used metrics for CTR prediction: 
\textbf{AUC} (Area Under ROC), higher AUC means better performance. 
Note that $0.1\%$ growth of AUC is considered as a significant 
improvement \cite{ref6} in the recommendation system. 
\textbf{LogLoss} (cross-entropy loss of the model), lower is better. 
To analyze memory costs of 
parameters in retrain stage, 
we also utilize 
\textbf{P-R}: the percentage of \textbf{P}arameter 
\textbf{R}eduction compared with UES-32.

\subsection{Overall Performance (RQ1)}
\label{sec.performance}
We now analyze the performance of various AES methods mentioned in Section.\ref{sec.details}. 
Note that we conduct AdaS\&S in two different modes: 
\textbf{AdaS\&S-R} is the resource-first mode, where larger penalty for resource constraint is applied; 
\textbf{AdaS\&S-E} is the effect-first mode, where searching stage will seek Emb-sizes with 
best recommendation accuracy regardless of resource competition penalty. 
The overall performance is shown in 
Table \ref{table.overall}. We can observe that: 
\begin{itemize}[%
  itemsep=0pt, partopsep=0pt, 
  parsep=\parskip, topsep=0pt,
  leftmargin=1em, listparindent=0em]
\item[1.] 
The impact of Emb-size is significant. 
It affects not only the effects of recommendation, 
but also the memory cost of model parameters. 
By comparing UES-32 and UES-24, reducing Emb-size from 32 to 24 will cause 
a decrease of $0.3\% \pm$ in AUC and an increase 
of $0.4\% \pm$ in LogLoss, with a parameter reduction $24\%+$. 

\item[2.] 


From the perspective of saving memory, 
MDE reduces the parameters by $20\%+$, 
but it cannot guarantee the effect of recommendation. 
Under the premise of keeping the recommendation effect unchanged, 
AutoDim and AutoEmb reduces the parameters by $10\%\pm$ (unstable), 
ESAPN and SSEDS reduces the parameters by $20\%$. 
Excitingly, AdaS\&S-R outperforms other AES Method with $30\%+$ 
parameters reduction and bring a little improvement of recommendation effect.

\item[3.] 
From the perspective of improving the recommendation effect, 
Past AES methods as well as AdaS\&S-R can improve AUC 
by roughly $0.1\% \pm$ and decrease LogLoss by $0.15\% \pm$. 
It is impressive that 
AdaS\&S-E can improve AUC by $0.3\% \pm$ and 
decrease LosLoss by $0.35\% \pm$, which considerably outperforms others.

\end{itemize}
To summarize, 
AdaS\&S-E can successfully improve the 
recommendation effect by above $0.3\%$ increase of AUC 
and reduce $20\% \pm$ of parameters. 
Meanwhile, AdaS\&S-R can significantly reduce $30\%+$ 
of parameters and improve the recommendation effect by about 
$0.1\%$ increase of AUC. 
AdaS\&S significantly outperforms AES methods in terms 
of effect and stability.

\subsection{Detailed Analysis}
\label{robust}
In this section, we further investigate characteristics of AdaS\&S to address several RQs 
discussed in Section.\ref{sec.experiment}.

\subsubsection{Robustness of Emb-size Searching (RQ2)}
\label{sec.robust1}
\begin{figure}[h!]
  \centering
  \includegraphics[width=1\linewidth]{
    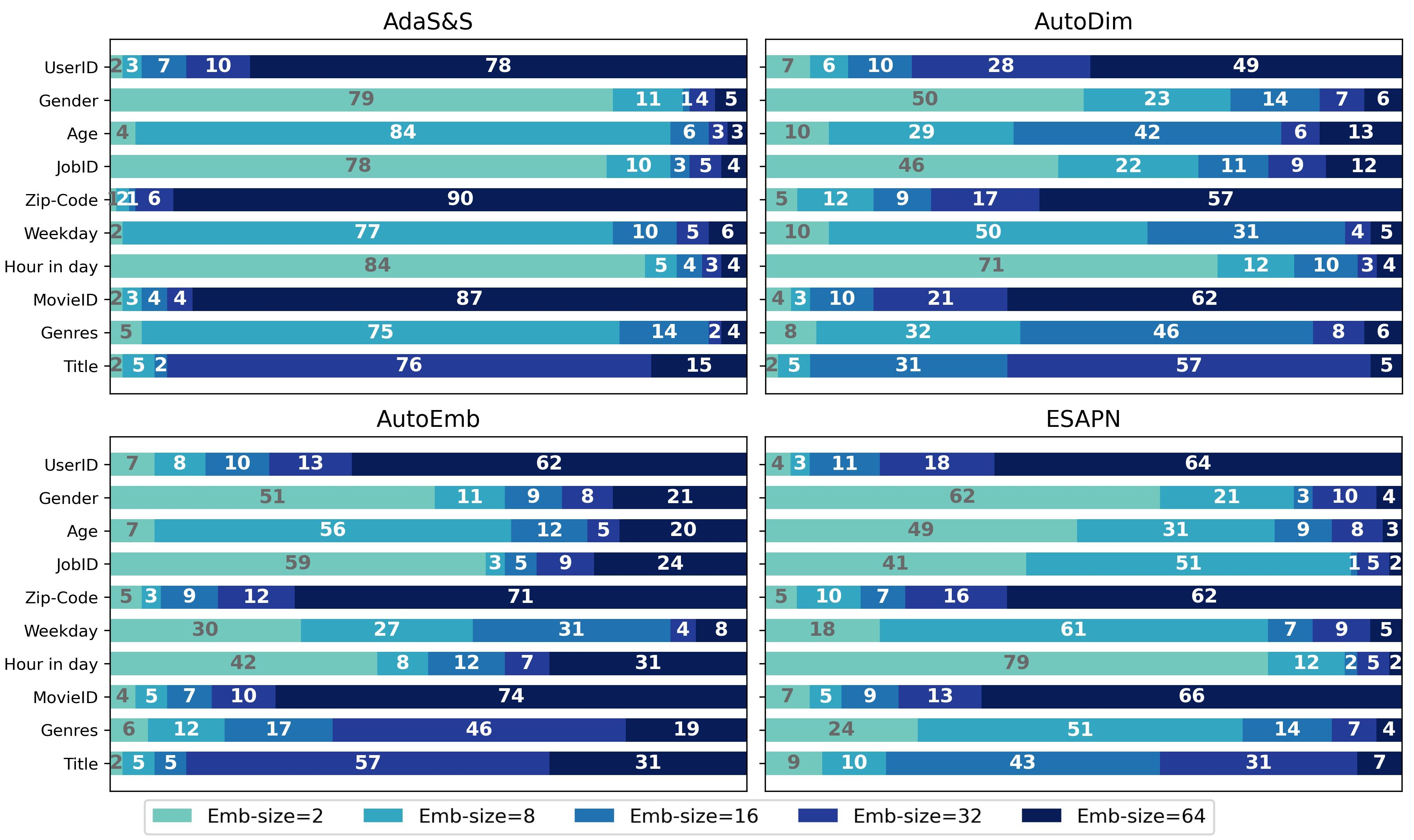
  }
  \caption{
    Stability of AES. Each row represents the number of 
    times a feature is assigned to various Emb-sizes in 100 
    searches. 
  }
  \Description{Stability}
  \label{fig.robust1}
  \vspace{-2pt}
\end{figure}

\quad\\
The decoupling of training and searching stage in AdaS\&S, as well as the design of 
Adaptive Sampling, is believed to greatly improve the stability of produced search results. 
To analyze the stability of searching,  
we conduct a series of experiments to investigate the 
distribution of selected Emb-sizes for features using AdaS\&S and other popular AES methods. 
For example, on ML-1M dataset we use 
different AES methods (AdaS\&S, AutoDim, AutoEmb, ESAPN) and 
perform 100 times of searching respectively. 
For each feature, 
if the Emb-size assigned to it is more consistent across these 100 results, 
then the corresponding AES method is believed to be more robust. 
We analyze the Emb-size searching results for 10 feature fields in ML-1M, 
and the result is shown in Figure \ref{fig.robust1}. 

In Figure \ref{fig.robust1}, for each feature 
AdaS\&S produces the same Emb-size with a probability of 
more than 75\%, sometimes even up to 90\%, indicating that 
the consistency of Emb-size across searching is considerably higher than AutoDim and other methods. 
In addition, we found that AutoEmb is more inclined to choose a 
larger Emb-size for all features, while ESAPN tends to choose smaller ones. 
In contrast, AdaS\&S does not have such issue, 
it assigns the most suitable 
Emb-size for each feature, e.g., the ``Gender'' feature is always set to smallest Emb-size 
and the ``MovieID'' always set to the largest. 
The analysis above demonstrates that AdaS\&S has better robustness compared 
with past AES methods, the Criteo and Avazu dataset also yields a similar conclusion.

\subsubsection{Ablation Study: Resources Competition Penalty (RQ3)}
\label{sec.robust2}
\begin{figure}[h!]
  \centering
  \includegraphics[width=1\linewidth]{
    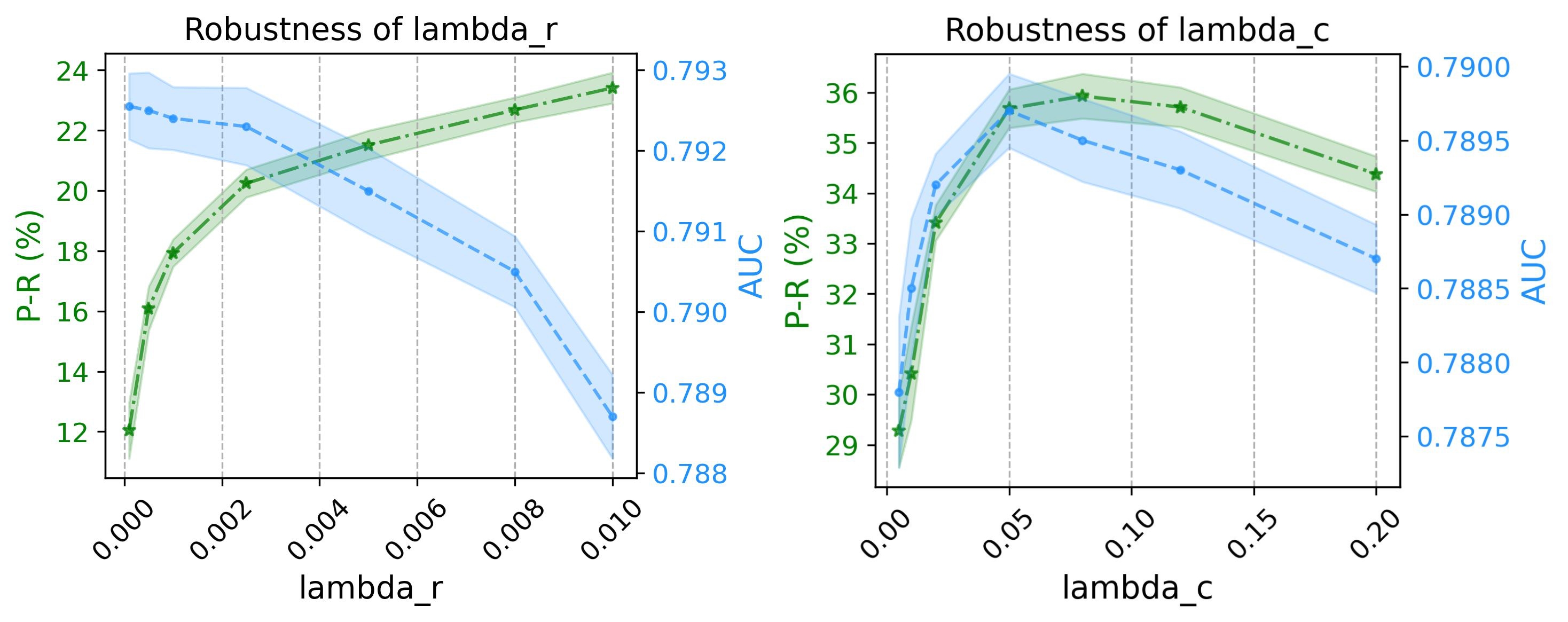
  }
  \caption{
    Effect of resources competition penalty. 
    P-R indicates the percentage of Parameters Reduce. 
    Left part is the robustness test of $\lambda_{r}$ 
    where $\lambda_{c}$ is set as 0.08. 
    Right part is the robustness test of $\lambda_{c}$ 
    where $\lambda_{r}$ is set as 0.005.
  }
  \Description{Stability}
  \label{fig.robust2}
  \vspace{-2pt}
\end{figure}

\quad\\
In order to demonstrate the effect of resource competition penalty in AdaS\&S, 
we conduct more experiments 
and analyze their results. 
With DeepFM as the DLRM, on ML-1M dataset we search the 
Emb-size assignment with different weights 
($\lambda_{r}$ and $\lambda_{c}$ in Eq.\eqref{eq.penalty}) 
in the resources competition penalty, 
then we retrain the model using the produced Emb-sizes. 
Performances of retrained models is shown in Fig.\ref{fig.robust2}. 
To obtain convincing results, all experiments are conducted for multiple times with 
various random seeds, the mean of metrics and standard error of 
the mean is reported. 

According to the left side of Fig.\ref{fig.robust2}, 
the larger $\lambda_{r}$ is, the more parameters is reduced. 
At the same time, the AUC metric drops and leads to worse recommendation 
effect. This indicates that 
larger $\lambda_{r}$ will result in stronger limitation on memory resources. 
The right side of Fig.\ref{fig.robust2} shows that, 
when $\lambda_{c} < 0.05$, AUC and P-R 
grows together with $\lambda_{c}$. When $\lambda_{c} > 0.05$, 
AUC will decline as $\lambda_{c}$ keeps increasing, 
and the growth of P-R also stopped after $\lambda_{c} > 0.08$. 
This indicates that 
larger $\lambda_{c}$ will result in fiercer competition between features. 
Features that are in a disadvantaged position in competition will be 
assigned a smaller Emb-size, yet it may need a larger Emb-size to fully 
represent the information within. 
Therefore, the recommendation effect will degrade if $\lambda_{c}$ is too large. 
In summary, the resource competition penalty (Eq.\eqref{eq.penalty}) can 
balance the computation resources 
and recommendation effectiveness during AES, guide the AES results to suit specific resource budget. 
In practice, $\lambda_{r}$ shall be adjusted manually 
based on actual requirements, and a moderate $\lambda_{c}$ is recommended.

\subsubsection{Ablation Study: Supernet Training (RQ4)}
\label{sec.robust3}
\quad\\
The consistency of the trained supernet is crucial for obtaining 
a proper and stable search results (Section.\ref{sec.preliminary.one-shot}). 
We utilize Kendall-tau correlation 
of architecture performance ranking based on subnet (inherited weights) and their ground truths 
(stand-alone trained weights) \cite{yu2019evaluating,zhang2020one}. Note that Kendall-tau value (dubbed as ``KD'') is in $[-1,1]$ and higher value indicates 
better supernet consistency. 
In particular, we randomly select 200 different subnet architectures (Emb-size allocations) 
to evaluate the KD value of AUC and LogLoss. 
The empirical results are shown in Table \ref{table.robust3}. 

\begin{table}[h!]
\vspace{2pt}
\renewcommand\arraystretch{1.1}
\centering
\caption{
  Consistency of supernet. 
}
\label{table.robust3}
\resizebox{.7\linewidth}{!}{%
\begin{tabular}{cccc}
  \Xhline{0.8pt}
  Scheme  & Sampling   & KD(LogLoss)  & KD(AUC) \\
  \Xhline{0.5pt}
  Shared  & Random     & 0.7246  & 0.7020 \\
  Shared  & Vanilla\&Uniform     & 0.7488  & 0.7351 \\
  Shared  & Weight\&Uniform   & 0.7541  & 0.7409 \\
  Shared  & Adaptive   & 0.7891  & 0.7815 \\
  Indep   & Random     & 0.7384  & 0.7412 \\
  Indep   & Vanilla\&Uniform   & 0.7498  & 0.7455 \\
  Indep   & Weight\&Uniform   & 0.7673  & 0.7547 \\
  Indep   & Adaptive   & \textbf{0.7904}  & \textbf{0.7970} \\
  \Xhline{0.8pt}
\end{tabular}%
}
\end{table}
In Table \ref{table.robust3}, 
``Indep'' is for the independent supernet scheme and ``Shared'' for 
the shared scheme. We also conduct ablation study on different training methods and compare the consistency: 
in the ``Sampling'' column, ``Random'' is the naive sampling method for training subnets 
similar to \cite{guo2020single}. At each training step, 
it selects a feature with the probability $p=0.6$, 
and then picks a candidate embedding of the feature with $p=1/T$.
The ``Vanilla\&Uniform'' method is from OptEmbed \cite{lyu2022optembed}, for each feature, 
it asummes the probability for training each dimension as a uniform distribution, i.e., 
sampling rate for candidate embeddings are proportional to their Emb-sizes. 
``Weight\&Uniform'' further selects features 
by weighted probabilities \cite{chu2021fairnas}, 
features with larger cardinality will have larger weight to be sampled, 
and the sampling for candidate Emb-sizes is same as ``Vanilla\&Uniform''. 
The ``Adaptive'' method is our proposed solution in Section.\ref{sec.supernet}. 

From Table \ref{table.robust3}, we can conclude that the ``Indep'' scheme performs 
better than ``Shared'', and we attribute this to the fact that there 
exists overlapping between different embedding tables in the ``Shared'' scheme. 
Note that given same sampling method, the ``Shared'' scheme is not 
much worse than ``Indep''. 
Further, 
the adaptive supernet training outperforms all other 
training methods. 
In summary, our adaptive supernet is of satisfactory 
consistency and could better guide the searching stage.

\subsubsection{Efficiency Analysis (RQ5)}
\label{sec.efficiency}
\begin{figure}[h!]
  \centering
  \includegraphics[width=0.9\linewidth]{
    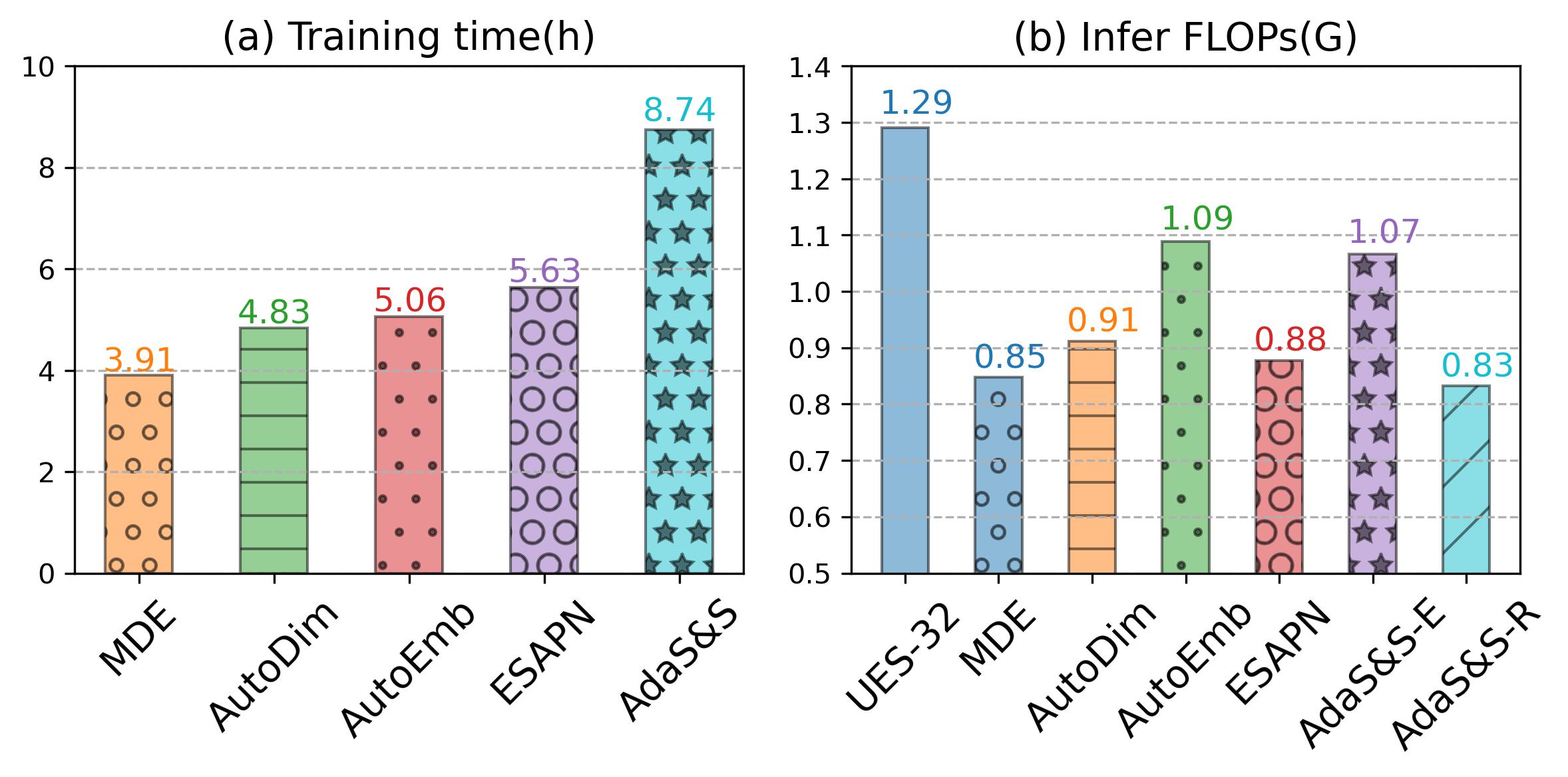
  }
  \caption{
    Training time and FLOPs of AES methods. 
  Left side shows training time of each AES methods. Right side compares FLOPs with unified 
  Emb-size models (UES-32).
  }
  \Description{Training time and FLOPs of AES methods. 
  We set the metric of UES-32 as 1.0 and show 
  the ratio of other methods compared to UES-32.}
  \label{fig.efficiency}
  \vspace{-2pt}
\end{figure}

\quad\\
Apart from model effectiveness, efficiency is also 
key consideration in real-world applications.
In this section, 
we investigate the influence of AdaS\&S on efficiency of AES and model inference. 
Specifically, we compare the training time of AES methods (without the model retraining) 
to show their searching efficiency. And we summarize 
Floating-point Operations (FLOPs) of the corresponding retrained model in searched Emb-sizes 
to indicate the inference efficiency. 
The experiment is conducted on Criteo dataset using Wide\&Deep architecture 
with one Tesla T4 GPU, we visualize the results in Figure \ref{fig.efficiency}. 

In Fig.\ref{fig.efficiency}(a), AdaS\&S requires highest 
training time to produce AES result, and we attribute this to the two-stage design of 
our framework, which inevitably add additional time costs (note that we show the 
total time 
cost of two stages together for AdaS\&S). It is worth mentioning that 
though total cost is higher, 
our approach benefits from the one-shot nature and the search stage is more 
light-weighted than supernet training stage, so 
the cost of AdaS\&S is much lower than others when conducting multiple times of 
searching for various requirements once the supernet is trained. 

From Fig.\ref{fig.efficiency}(b), AES methods could improve the inference efficiency because of 
the parameter reduction (illustrated in Table \ref{table.overall}). 
Despite using the effect-first mode, AdaS\&S-E could still 
achieve FLOPs comparable to other methods. 
The resource-first mode (AdaS\&S-R) significantly reduce the inference 
overhead with lowest FLOPs. 
In summary, the superior efficiency of AdaS\&S makes it 
quite suitable to be deployed in real-world tasks.


\section{RELATED WORK}
\textbf{AutoML Techniques:}  
As a core component of AutoML, Neural Architecture Search(NAS) has raise research 
interests since \cite{ref17}. 
To address the computation overhead, some researches \cite{ref22,zheng2023automlsurvey,ref4} 
utilize parameter sharing to search sub-graphs from a large network graph. 
Bender et al. \cite{bender2018understanding}  systematically 
formulate the supernet method and identify the effectiveness of weight sharing. 
Guo et al. \cite{guo2020single} improve the supernet optimization with uniform path sampling, 
FairNAS \cite{chu2021fairnas} further 
propose 
the strict fairness sampling method. 
Other studies \cite{ref18,chen2019progressive,xie2018snas,ref18} model the 
network architecture into continuous search space. With differentiable searching, 
these method can optimize with gradient descent. 

\hspace{-1em}\textbf{AES methods:} 
Heuristic, NAS-based, and pruning methods are primary categories of AES methods. 
As a heuristic method, 
MDE \cite{ginart2021mixed} set Emb-sizes based on popularity of features. However, 
these rule-based 
solutions are not able to incorporate with task-specific models, 
and the performance is not guaranteed \cite{yan2021learning}.
Recent studies \cite{autodim,ref19,liu2020automated,zhaok2021autoemb,ref20} 
introduce NAS into AES. 
AutoEmb \cite{zhaok2021autoemb} selects Emb-sizes with a controller network 
and update it together with main network 
by bi-level optimization in DARTS \cite{ref18}. 
AutoDim \cite{autodim} utilize the DARTS framework to optimize weights 
over different Emb-sizes. 
NIS \cite{ref19} searches the optimal 
embeddings for each feature value in a discrete space, and parameters for embeddings and controller 
are optimized jointly. 
These methods can hardly guide searching results and meet specific resource demands, and 
the jointly optimizing 
of embedding training and searching still needs further study. 
Other methods would prune embedding parameters to smaller 
Emb-sizes. 
For example, 
AMTL \cite{yan2021learning} introduces a twins-based layer 
to generate mask 
vectors to prune embeddings. 
PEP \cite{liu2021learnable} removes redundant dimensions 
with the help of a learnable threshold for embedding values. 
SSEDS \cite{qu2022single} prunes Emb-sizes by measuring 
the influence of each dimensions on the loss. 
The one-shot supernet for embeddings in OptEmbed \cite{lyu2022optembed} is similar with ours, yet it 
trains supernet with uniform sampling and searches dimension masks by Evolution Algorithm. 
The learning of pruning thresholds or masks in AES methods above raises the risk of unstable 
searching results, 
which also undermines the interpretability of outcomes.


\section{CONCLUSION}
In this paper, we propose a novel automatic embedded size search (AES) 
method called AdaS\&S. 
Specifically, we design a one-shot supernet framework 
to tackle the AES problem. 
First, we decouple main model training from the search process, 
which ensures the soundness of parameters and 
stability of Emb-size allocation. 
We design the Adaptive Sampling 
method for training supernet, which greatly improves its consistency. 
Second, in the searching stage, we devise a RL-based method to 
seek the optimal AES result. 
Finally, a well-designed resource competition penalty is employed to 
balance the performance and the resources cost. 
We evaluate the AdaS\&S framework with extensive experiments 
on widely used datasets, showing that 
our framework can outperform other AES methods and address the key challenges 
in past research.
The source code of the proposed framework is available 
online\footnote{\url{https://github.com/}XXXXX/XXXXX/}. 

In the future, we would like to further study the training efficiency of AdaS\&S 
and try to reduce the total time cost needed for two stages. 
We will also explore more ways to leverage the prior knowledge about the 
recommendation tasks and features to improve the model effectiveness. 











\end{document}